\documentstyle[12pt,openbib,epsf,axodraw]{article}

\input epsf
\def\br{\begin{eqnarray}}
\def\er{\end{eqnarray}}
\def\be{\begin{equation}}
\def\ee{\end{equation}}
\def\l{\label}
\def\a{\alpha}
\def\b{\beta}

\def\L{\Lambda}

\def\G{\Gamma}

\def\m{\mu}
\def\n{\nu}
\def\P{\Pi}
\def\ra{\rightarrow}
\def\d{\delta}
\def\<{\left\langle}
\def\>{\right\rangle}
\begin{document}
\vspace{1cm}
\begin{center}
{\large\bf Infrared finite solutions for the gluon propagator
and the QCD vacuum energy}\\[.4cm]       
J.~C.~Montero~\footnotemark
\footnotetext{e-mail: montero@axp.ift.unesp.br}, 
A.~A.~Natale~\footnotemark
\footnotetext{e-mail: natale@axp.ift.unesp.br} 
and P.~S.~Rodrigues da Silva~\footnotemark\\
\footnotetext{e-mail: fedel@axp.ift.unesp.br}
Instituto de F\'{\i}sica Te\'orica,                                         
Universidade Estadual Paulista\\
Rua Pamplona, 145, 01405-900, Sao Paulo, SP, Brazil
\end{center}
\thispagestyle{empty}
\vspace{1cm}

\begin{abstract}    
Nonperturbative infrared finite solutions for the gluon 
polarization tensor have been found, and the
possibility that gluons may have a dynamically
generated mass is supported by recent Monte Carlo simulation
on the lattice. These solutions differ among themselves, due
to different approximations performed when solving the Schwinger-Dyson 
equations for the gluon polarization tensor. 
Only approximations that minimize energy are
meaningful, and, according to this, we compute an effective potential for 
composite operators as a function of these solutions in order
to distinguish which one is selected by the
vacuum. 
\end{abstract} 
\newpage

During the last years there has been much effort in trying to
obtain a non-perturbative form for the gluon 
propagator~[1-8], and perhaps one of the most interesting
results is the one where it is argued that the gluon
may have a dynamically generated mass~\cite{corn}. The
existence of a mass scale, or the absence of a pole
at $k^2=0$ is natural if one assumes that gluons do
not propagate to infinity, {\it i.e.} these propagators
describe confined gluons. From
the phenomenological point of view, this possibility
may shed light on several reactions where long distance
QCD effects can interfere, and examples about the consequences
of massive gluons can be found in the literature, see,
for instance, Ref.[9-11].

The study of the infrared behavior of the gluon propagator
was also performed numerically~\cite{early}, and 
the last numerical lattice computations give strong 
evidence for an infrared finite gluon propagator~\cite{now}.
The most recent results~\cite{now} seems 
to exclude a propagator of the form $1/k^4$ determined
in several analysis of Schwinger-Dyson equations (SDE) and
claimed to give confinement~\cite{mand,baker,brown}.
In particular,  the results of Bernard {\it et 
al.}~\cite{now} are consistent with an infrared
finite propagator, but can be fitted by the
Cornwall's massive propagator~\cite{corn} as well as by
the one found by Stingl {\it et al.}~\cite{sting}.
These lattice computations could be criticized because of
the gluon propagator gauge dependence. However, there is
also a lattice study of the ``gauge-invariant" two-point
correlation function of the gauge field strengths by
Di Giacomo {\it et al.}~\cite{giacomo}, where a finite correlation
length for the gluons in the QCD vacuum is found, indicating
the presence of a mass scale for the gluon propagation. 

If the lattice results can be considered exciting,
the same cannot be said about the analytic calculations.
Looking at Ref.~\cite{mand} to \cite{lav} we verify that
infinite as well as infrared finite non-perturbative
propagators have been found in
the literature, although this difference can be credited
to the fact that some authors discarded from the
beginning an infrared finite solution for the gluon
propagator. Even if we separate the solutions of
SDE  
for the gluon self-energy, according
to the existence or not of a pole at $k^2 = 0$, we still
remain with different forms of non-perturbative propagators,
which reflect the different approximations 
involved in the calculations, for instance,
different truncations of the integral equations,
or the handling of higher-point functions which
are present in the two-point one, {\it etc...}

In this work we propose to compute the effective potential
for composite operators~\cite{cornja} at stationary
points as one method to distinguish among the different
solutions, {\it i.e.} we will calculate the vacuum energy for
some infrared finite non-perturbative Schwinger-Dyson
solutions of the gluon propagator, and the vacuum will
select the solution that leads to the deepest minimum
of energy. Therefore, this will be a complementary
tool to find solutions of the gluonic SDE, {\it i.e.} they 
should be tested by means of
the effective potential calculation.
As we are far from being able
to solve the SDE without any simplification,
only approximations that minimize the
vacuum energy should be performed when solving these
integral equations. We will also discuss how the
SDE solutions can be constrained phenomenologically.

For a non-abelian gauge theory the effective potential
has the form~\cite{cornja}              
\br
V(S,D,G) &=& - \imath \int \frac{d^4p}{(2\pi)^4} 
Tr ( \ln S_0^{-1}S - S_0^{-1}S + 1) \nonumber \\
&& - \imath \int \frac{d^4p}{(2\pi)^4}
Tr ( \ln G_0^{-1}G - G_0^{-1}G + 1) \nonumber \\
&& + \frac{1}{2} \imath \int \frac{d^4p}{(2\pi)^4}
Tr ( \ln D_0^{-1}D - D_0^{-1}D + 1) \nonumber \\
&& + V_2(S,D,G), 	
\l{vfull}                                                              
\er                                                        
where $S$, $D$ and $G$ are the complete propagators of
respectively fermions, gauge bosons and Faddeev-Popov
ghosts; $S_0$, $D_0$ and $G_0$ the respective bare
propagators. $V_2(S,D,G)$ is the sum of all two-particle
irreducible vacuum diagrams,  depicted in Fig.(1), and 
the solutions of the equations   
\be
\frac{\d V}{\d S}=\frac{\d V}{\d D}=\frac{\d V}{\d G}=0,
\l{delv}
\ee
give the SDE for fermions, gauge
bosons and ghosts.
                            
We can represent $V_2(S,D,G)$ analytically by
\br
\imath V_2(S,D,G) &=& - \frac{1}{2} Tr(\G S \G S D)
- \frac{1}{2} Tr (F G F G D) \nonumber \\
&& + \frac{1}{6} Tr(\G^{(3)}D\G^{(3)}DD)
+ \frac{1}{8} Tr(\G^{(4)}DD)
\l{v2full}
\er
where $\G$, $F$, $\G^{(3)}$ and $\G^{(4)}$ are respectively
the proper vertex of fermions, ghosts, trilinear and quartic
gauge boson couplings~\cite{natale}. In Eq.(~\ref{v2full}) we 
have not written the gauge and Lorentz indices, as well as the
momentum integrals.        

The complete gauge boson propagator $D$ is related
to the free propagator by
\be
D^{-1} = D_0^{-1} - \Pi,
\l{dfull}
\ee
where $\Pi$ is the gluon polarization tensor,
which is obtained from Eq.(\ref{vfull}) and Eq.(\ref{delv}), 
and described by
\be
\Pi = \G S \G S + F G F G 
- \frac{1}{2} \G^{(3)} D \G^{(3)} D - \frac{1}{2} \G^{(4)} D.
\l{picomp}
\ee
The diagrams contributing to $\Pi$
are shown in Fig.(2), and this
self-energy is the one that has been solved in
many different ways, leading to a series of
non-perturbative forms for the gluon propagator
in the infrared region~[1-6]. As our intention is to compute
the effective potential through a variational
approach, the best ansatz for the complete
propagator will be given by the 
approximate solution
of $\Pi$.
                                         
The physically meaningful quantity we need to compute
is the vacuum energy density given by
\be
\Omega= V(S,D,G) - V(S_p,D_p,G_p),
\l{omega}
\ee
where $D_p$ is the perturbative
counterpart of $D$, and
we are subtracting from the potential the 
perturbative part that does not contribute to dynamical 
mass generation, denoted by $V(S_p,D_p,G_p)$. As our 
interest is concentrated in the pure glue theory,
we disregard all the quark contributions for the
effective potential from now on, {\it i.e.} we
drop the first contribution on the right-hand side
of Eq.(~\ref{v2full}). It is opportune to recall
that for perturbative gluons $\Omega$ is plagued
by infrared divergences, and only for the infrared
finite gluon propagators it can be
calculated without ambiguities~\cite{corn}.
                   
Before we start calculating $\Omega$ it is convenient
to discuss how the solutions of $\Pi$ are
determined. One of the first calculations of the
gluon polarization tensor is due to Mandelstam~\cite{mand},
where the ghosts and the diagram with quartic
coupling are disregarded. The neglect of ghosts
diagrams was shown to be correct, because their
contribution is numerically small~\cite{mand}.
These approximations, with the use of the
Landau gauge, were shown to be satisfactory in 
the lengthy and detailed work of
Brown and Pennington~\cite{brown}. With this
approximation only the diagram with trilinear
coupling should be considered in Eq.(~\ref{v2full}),
{\it i.e.} the gluon polarization tensor is
going to be given by
\be
\Pi = - \frac{1}{2} \G^{(3)} D \G^{(3)} D.
\l{piapro}
\ee
More complex approximations can be
used, but in practice only
the diagram with the trilinear coupling has been 
considered in most of the works up to now. As far as
we know, few papers have gone beyond these 
approximations, and one of these
approaches is the Cornwall's determination,
through the use of the pinch-technique,
of a set of diagrams leading to a gauge invariant
SDE~\cite{corn}. Another one is
the Stingl {\it et al.}  manipulation of the 
higher-point functions~\cite{sting}, although
this calculation was done in Landau gauge.

The vacuum energy can be computed with the same
approximations performed to obtain solutions
of the gluon self-energy. However, a
much better approximation to study the vacuum
stability results when we plug the stationary conditions
(Eq.(~\ref{delv}) and Eq.(\ref{dfull})) into $\Omega$,
obtaining an expression for the values of $\Omega$ at
its minimum, which will be denoted by $\< \Omega \>$.
For example, with the Mandelstam approximation the
2PI diagrams contributing to $\Omega$ would be given
by
\be 
\Omega_2 = \frac{-\imath}{6} Tr(\G^{(3)}D\G^{(3)}DD)-
(D\rightarrow D_p),
\l{ommand}
\ee
but applying the stationary condition Eq.(\ref{delv})
we obtain 
\be
\< \Omega_2 \> = \frac{\imath}{3} Tr(\Pi D) - (D\rightarrow D_p) ,
\l{ommin2}
\ee
where the
calculation is now reduced to an ``one-loop" integration.
It is important to stress that $\< \Omega \>$ is much
less dependent on the ansatze we use to compute
the vacuum energy. This has already been emphasized
in calculations of the vacuum energy for chiral
symmetry breaking~\cite{castorina}. Note that 
$\< \Omega \>$ also does not suffer from possible
problems in the determination of the effective potential
for composite operators~\cite{banks}.

The vacuum energy at stationary points, within the
approximation discussed above, is given by
\br
\< \Omega \> &=& \frac{\imath}{2} Tr [ \ln (D_0^{-1}D)
- D_0^{-1} D + 1 ] \nonumber \\
&& - \frac{\imath}{3} Tr [(D^{-1} - D_0^{-1}) (D - D_0)],
\l{finome}
\er
where we assumed $D_p=D_0$.

Displaying the momentum integration, using
the Landau gauge expression for the complete
gluon propagator
\be
D^{\m \n}(p^2) = - \frac{\imath}{p^2-\P (p^2)} 
( g^{\m \n} - \frac{p^\m p^\n}{p^2} ),
\l{dlprop}
\ee
and calculating the trace,
Eq.(\ref{finome}) can be cast in the following form
\be
\< \Omega \> = - \frac{3(N^2 -1)}{2} \int \frac{d^4P}{(2\pi)^4} \,
\left[ \frac{\Pi}{P^2+\Pi} - \ln ( 1+\frac{\Pi}{P^2} ) +
\frac{2}{3} \frac{\Pi^2}{P^2(P^2+\Pi)} \right],
\l{ommom}
\ee
where all the quantities are in Euclidean space
$(P^2=-p^2)$, and $N=3$
for QCD. Note
that the computation of the full vacuum energy $\Omega$
requires the use of nonperturbative vertices satisfying
the Slavnov-Taylor identities, which give strong
relations between the diagrams of Fig.(1)~\cite{corn2}, and
we escaped from this difficulty in our $\< \Omega \>$
calculation.

To compute $\< \Omega \>$ we need only a reasonable ansatz
for the gluon propagator or its self-energy $\Pi$,
which will be given by the SDE solutions to be
presented in the sequence. One of the 
infrared finite propagators found in the
literature was determined by Cornwall~\cite{corn}
\be
D_c^{-1}(P^2) = [P^2 + m_g^2(P^2)]bg_s^2 \ln [
\frac{P^2 +4 m_g^2(P^2)}{\L^2}],
\l{dcorn}
\ee
where $m_g(P^2)$ is the dynamical gluon mass 
\be
m_g^2(P^2)=m_g^2 \left[ {\ln\left(
\frac{P^2+4m_g^2}{\L^2} \right)}/
{\ln{\frac{4m_g^2}{\L^2}}} \right]^{-12/11},
\label{mcor}
\ee
$m_g$ is the gluon mass scale, $g_s $ the strong coupling constant, 
$b=33/48\pi^2$ is the leading order coefficient
of the $\b$ function of the renormalization group equation,
and $\L =  \L_{QCD}$ is the QCD scale. This form for the
propagator was obtained as a fit to the numerical solution
of a gauge invariant set of diagrams for the gluonic SDE,
which also imposes $g_s \simeq 1.5 - 2$.

Eq.(\ref{dcorn}) is consistent with the expected
ultraviolet behavior of the gluon propagator
\be
D_{UV}^{-1}(P^2) \ra P^2 \ln (P^2). 
\l{uvpro}
\ee
However, the dynamical gluon mass $m_g(P^2)$ is
expected to decrease 
faster asymptotically as predicted by the
operator product expansion 
(OPE)~\cite{cornhou,lav}. It should be remembered
that the SDE for the gluon propagator
is accurate in the determination of the infrared
dynamical mass, but its ultraviolet behavior can
be better determined by means of the OPE.
The behavior of the dynamical gluon mass
through the OPE was determined by Lavelle~\cite{lav},
and is equal to
\be
\Pi_{UV} (P^2) \sim - \frac{34 N \pi}{9(N^2-1)}
\frac{\< \a_s G^2 \>}{P^2},  
\l{piuv}
\ee
where $\< \a_s G^2 \>$ is the gluon condensate.
Note that as long as the gluon self-energy 
is infrared finite, and no matter 
its ultraviolet behavior is given by Eq.(\ref{mcor})
or Eq.(\ref{piuv}), it is
easy to see that $\< \Omega \>$ is free of 
divergences~\cite{corn}.

Another infrared finite propagator has been found by Stingl
{\it et al.}~\cite{sting}, its form agrees with that
derived by Zwanziger based on considerations related to
the Gribov horizon~\cite{grib}, and is given by
\be
D_s(P^2)= \frac{1}{P^2 + \frac{\m_s^4}{(P^2 + c^2)}},
\l{dstin}
\ee
where $\m_s$ is a mass scale not determined in Ref.~\cite{sting}.
Eq.(~\ref{dstin}) with $c=0$ is the one found by 
Zwanziger~\cite{grib}. As claimed in the last paper
of Ref.~\cite{sting}, the full propagator should contain
a $c \neq 0$ term, however, due to the complexity introduced
by such factor, this case was not studied in
detail. The term $\m_s^4/P^2$ is exactly what is expected
by OPE analysis whenever a mass scale for the gluon is
introduced, as shown by Eq.(~\ref{piuv}), therefore,
its appearance is not surprising. With $c=0$ the
propagator goes to zero at $P^2=0$, however the
gluon self-energy diverges, and we discard this
possibility here assuming that $c=\m_s$, which
could be expected as a natural mass scale 
leading to a finite gluon polarization tensor.
The Bernard {\it et al.} 
lattice result for the gluon 
propagator~\cite{now} can be fitted by Eq.(\ref{dcorn})
as well as Eq.(\ref{dstin}).

Finally, Marenzoni {\it et al.}~\cite{now} also performed a
lattice study of the gluon propagator in the Landau gauge,
obtaining for its infrared behavior the following fit
\be
D_m(K^2) = \frac{1}{m^2_l + Z P^2 (\frac{P^2}{\L^2})^{\eta}},
\l{mar}
\ee
where $m_l$, $Z$ and $\eta$ are constants determined with
the numerical simulation. $m_l$ is 
of ${\cal O}(\L \simeq 160~MeV)$,
$Z \simeq 0.4$, and $\eta \simeq 0.5$
what is slightly different from the previous
propagators. The results of Bernard {\it et al.} also
show the behavior $(P^2)^{\eta}$, but with a smaller
value for $\eta$.
We will consider only the propagators discussed up
to now. Although other propagators can still be found in the 
literature, the propagators that we presented above
are the most discussed ones and enough
for our purposes.

The result of Ref.~\cite{corn}, which led
to Eq.(\ref{dcorn}), was
shown to be gauge invariant, and the others were obtained
in the Landau gauge, which we assume here expecting that
any gauge dependence should be negligible.            
In the above equations for the infrared propagator, 
except from Cornwall's one, it
is not clear how to connect these expressions with their
ultraviolet part, and this is an extra complication,
because any, {\it ad hoc}, link between the ultraviolet
and infrared propagator will introduce a new mass scale.
It is obvious that the physics should not depend on this
particular scale, but it does introduce some arbitrariness.
We, therefore, will not consider the 
ultraviolet renormalization
group logarithmic behavior of the full propagator,
as described by Eq.(\ref{uvpro}), since it should be common
to all propagators (although it is only present in
Eq.(\ref{dcorn})). In the especific case of the lattice propagator, 
Eq.(\ref{mar}), in which there is no clear separation between the
infrared and ultraviolet regions, we shall simply neglect its
peculiar form, assuming that the vacuum gluon polarization can be
cast as a pole in the perturbative propagator factorized from 
Eq.(\ref{mar}).

We adopt the following ansatze to 
calculate $\< \Omega \>$: a) $\P (P^2) = m_g^2 (P^2)$,
where $m_g^2 (P^2)$ is given by Eq.(\ref{mcor}); 
b) To be consistent with the 
massive Cornwall propagator and with OPE, we will consider
a gluon self-energy that interpolates between the constant
infrared behavior of Eq.(\ref{mcor}) and the 
ultraviolet one of Eq.(\ref{piuv}), which is
given by
\be
\P (P^2)= \m^2 \theta(\m^2-P^2) + \frac{\m^4}{P^2}
\theta(P^2-\m^2);
\l{pian}
\ee
c) $\P (P^2)$ consistent with Eq.(\ref{dstin}), {\it i.e.}
\be
\P (P^2) = \frac{\m_s^4}{P^2 + \m_s^2};
\l{pstin}
\ee
d) We assume that $\P (P^2)$
for the lattice inspired propagator is given by 
\be
\P (P^2) = m_l^2 \theta(\m^2-P^2) + 
\frac{m_l^2}{Z(P^2/\Lambda^2 )^\eta} \theta(P^2-\m^2),
\l{pmaren}
\ee
which follows from considering the generated gluon mass
as a pole in the perturbative propagator, factorized 
from Eq.(\ref{mar}).
Finally, since $m_g$ and $\m_s$ are not determined
by the solution of the SDE, we  
compute the vacuum energy as a function
of these parameters. 

In Fig.(3) we show the vacuum energy computed
with our ansatze a) to d), as a function of
the gluon mass scale ($m_g \equiv \m_s \equiv m_l$). 
Note that once selected a form for the gluon self-energy,
$\Pi$, the $\<\Omega\>$ calculations showed in 
Fig.~(\ref{fig:3}) only give us the vacuum energy as a
function of the gluon mass scale (or $m_g/\Lambda$),
and we do not have a criterion to fix a particular value
of this scale parameter. We can only compare the different
vacuum energy curves, corresponding to the different 
gluon self-energy behavior, assuming a common 
value for the gluon mass scale.
From the behavior of the vaccum energy presented in this
figure we verify that
the Cornwall propagator, Eq.(\ref{dcorn}),
leads to the deepest values of minimum of the vaccum energy,
as long as $m_g > 1.6 \L$, 
compelling us to believe that this could be the most favoured
form of the gluon propagator to match the one chosen by
the QCD vacuum, for gluon masses obeying such relation. For
smaller gluon masses ($m_g < 1.6 \L$), the lattice propagator
is the one that best represents the actual gluon propagator. 
It is clear that                     
apart from the lattice propagator (Eq.(\ref{mar})),
where all the constants are determined, we can only compare
the different values for the vacuum energy, if we are
able to constrain the parameters $m_g, \m_s $.
In principle they should be 
related to the scale $\L$, but we are far from this
achievement, and we discuss in the sequence
some of the constraints that have been put
forward on these parameters.

The phenomenological constraints on the dynamical
gluon mass may come from the quarkonium decays into
gluons~\cite{parisi,field}, potential models for
glueballs~\cite{soni}, approximations to the
QCD-Hamiltonian combined with a variational method~\cite{ji},
a possible relation between the gluon condensate and the
gluon mass through QCD sum-rules~\cite{graziani}, the
two-gluon exchange contribution to the Pomeron~\cite{hal},
and the vacuum functional of SU(3) Yang-Mills theory~\cite{kogan}.
The gluon masses obtained in most of these studies are
of the order of $2 \L$, but one should be careful when
considering this value because with few exceptions these
calculations consider only a bare gluon mass, and the
specific behavior of the dynamical mass with the momentum 
may modify the result. These calculations in general also involve
many different (and some untenable) approximations.
Therefore, we are in the safe side if we consider that the gluon
mass may lay between $\L$ and $3 \L$, and according to
Fig.(3), the propagator determined by Cornwall and the
one obtained in the lattice simulation lead to the deepest
vacuum energy, and approach the actual behavior of the
confined gluon propagator. Finally,
we can also use one relation found by Cornwall~\cite{corn}
between the gluon condensate and the vacuum energy 

\be
\< g^2 G^a_{\m \n} G^{\m \n}_a \> = - 8 b^{-1} 
\< \Omega \>,
\l{cond}
\ee
which allows a determination of the gluon mass. To arrive
at this relation it has been assumed that the $\b$ function
is given by its perturbative behavior ($\b \simeq -bg^3$).
Although the $\b$ function
can be different from this in the condensed phase, maybe this
relation is the most trustable one to allow a determination
of the gluon mass. The left-hand side of Eq.(\ref{cond}) has
been evaluated through QCD sum-rules~\cite{shif} and is 
equal to $0.47$ GeV$^4$. Therefore, assuming $\L = 0.3$ GeV
we obtain the following gluon masses in the cases we
discussed above: a) $ 527$ MeV, b) $627$ MeV, c)
$715$ MeV, and d) $560$ MeV. Considering all the
gluon mass determinations in the literature we verify again
that the propagators of Ref.~\cite{corn} and Ref.~\cite{now}
are the ones to provide the most consistent picture.

In conclusion, starting from the effective action for
composite operators, we determined an expression for the
vacuum energy of QCD without fermions in the Landau gauge. 
The vacuum energy
calculation was used as a criterion to determine which, among 
a series of different forms determined for the gluon propagator, 
leads to the deepest minimum of energy. We verify that two
expressions are clearly favored and, consistently, when
we compare the vacuum energy with the gluon condensate, we
obtain gluon masses for these propagators that are in the
range of several determinations of this mass existent
in the literature.

\section*{Acknowledgments}                    

This research was supported in part by the
Conselho Nacional de Desenvolvimento
Cient\'{\i}fico e Tecnol\'ogico (CNPq) (JCM and AAN), and in
part by Funda\c{c}\~ao de Amparo \`a Pesquisa do Estado de 
S\~ao Paulo (FAPESP) (PSRS).

\newpage
\begin{figure} 
\vskip 1in
\centerline{
\begin{picture}(300,100)(-15,-15)
\put(-35,105){\makebox(0,0)[br]{$V_2(S,D,G)\;\; =$}}
\put(-5,105){\makebox(0,0)[br]{$-\frac{1}{2}$}}
\Gluon(0,110)(75,110){3}{8}
\ArrowArc(37.5,110)(37.5,0,180) 
\ArrowArc(37.5,110)(37.5,180,360) 
\put(145,105){\makebox(0,0)[br]{$-\frac{1}{2}$}}
\Gluon(150,110)(225,110){3}{8}
\DashCArc(187.5,110)(37.5,0,180){3} 
\DashCArc(187.5,110)(37.5,180,360){3} 
\put(-5,0){\makebox(0,0)[br]{$+\frac{1}{6}$}}
\Gluon(0,5)(75,5){3}{8}
\GlueArc(37.5,5)(37.5,0,180){3}{8} 
\GlueArc(37.5,5)(37.5,180,360){3}{8} 
\put(145,0){\makebox(0,0)[br]{$+\frac{1}{8}$}}
\GlueArc(187.5,23.75)(18.75,-90,270){3}{15}
\GlueArc(187.5,-13.75)(18.75,90,450){3}{15}
\end{picture}
}
\label{fig:1}
\vskip 0.5in
\caption{Two-particle irreducible vacuum diagrams contributing
to the effective potential.}
\end{figure}
\begin{figure} 
\vskip 1in
\centerline{
\begin{picture}(300,100)(-15,-15)
\put(-35,105){\makebox(0,0)[br]{$\Pi\;\; =$}}
\Gluon(-20,110)(10,110){3}{4}
\Gluon(65,110)(95,110){3}{4}
\ArrowArc(37.5,110)(27.5,0,180) 
\ArrowArc(37.5,110)(27.5,180,360) 
\Gluon(130,110)(160,110){3}{4}
\Gluon(215,110)(245,110){3}{4}
\DashCArc(187.5,110)(27.5,0,180){3} 
\DashCArc(187.5,110)(27.5,180,360){3} 
\put(-25,0){\makebox(0,0)[br]{$-\frac{1}{2}$}}
\Gluon(-20,5)(10,5){3}{4}
\Gluon(65,5)(95,5){3}{4}
\GlueArc(37.5,5)(27.5,0,180){3}{8} 
\GlueArc(37.5,5)(27.5,180,360){3}{8} 
\put(125,0){\makebox(0,0)[br]{$-\frac{1}{2}$}}
\GlueArc(187.5,23.75)(18.75,-90,270){3}{15}
\Gluon(130,5)(187.5,5){3}{7}
\Gluon(187.5,5)(245,5){3}{7}
\end{picture}
}
\label{fig:2}
\vskip 0.5in
\caption{Diagrams contributing to the gluon polarization tensor.}
\end{figure}
\newpage
\begin{figure}[htb]
\epsfxsize=0.8\textwidth
\begin{center}
\leavevmode
\epsfbox{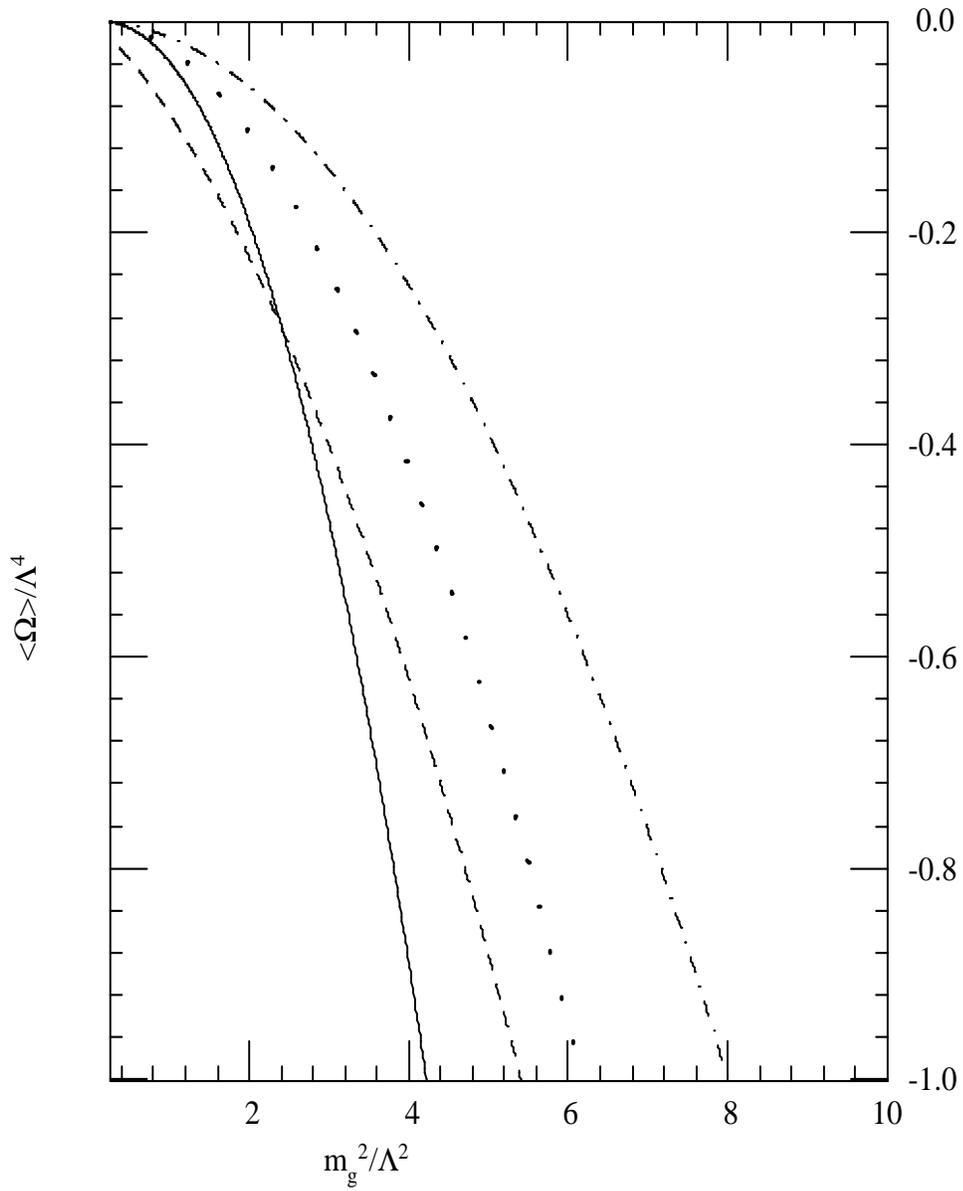}
\end{center}
\caption{Result of the vacuum energy calculation as a function
of the gluon mass for different ansatze for the gluon vacuum polarization.
The curves are related to the ansatze assumed inside the text:  
a) solid line Eq.(\ref{mcor}); b) dotted line Eq.(\ref{pian});
c) dashed-dotted line Eq.(\ref{pstin}); d) dashed line Eq.(\ref{pmaren}).}
\label{fig:3}
\end{figure}
\end{document}